\documentclass[showpacs,aps,prb,twocolumn,superscriptaddress,preprintnumbers,showkeys,amsmath,amssymb,floatfix,10pt]{revtex4-1}

\usepackage[utf8x]{inputenc}
\usepackage{graphicx}
\usepackage{dcolumn}
\usepackage{bm}

\begin{document}
\title{One dimensional Si-in-Si(001) template for single-atom wire growth}

\author{J.H.G. Owen}
\author{F. Bianco}
\author{S. A. K\"oster}
\author{D. Mazur}
\affiliation{University of Geneva, Department of Condensed Matter Physics,
NCCR MaNEP, 24 Quai Ernest-Ansermet, 1211 Geneva 4, Switzerland}

\author{D.R. Bowler}
\affiliation{University College London, Department of Physics \& Astronomy, Gower St, London WC1E~6BT, UK}
\affiliation{London Centre for Nanotechnology, 17--19 Gordon St, London WC1H~0AH, UK}

\author{Ch. Renner}
\affiliation{University of Geneva, Department of Condensed Matter Physics,
NCCR MaNEP, 24 Quai Ernest-Ansermet, 1211 Geneva 4, Switzerland}
\email{christoph.renner@unige.ch}

\date{\today}

\begin{abstract}
Single atom metallic wires of arbitrary length are of immense technological and
scientific interest. We describe a novel silicon-only template enabling the
self-organised growth of isolated micrometer long surface and subsurface
single-atom chains. It consists of a one dimensional, defect-free reconstruction
-- the Haiku core, here revealed for the first time in details -- self-assembled
on hydrogenated Si(001) terraces, independent of any step edges. We discuss the
potential of this Si-in-Si template as an appealing alternative to vicinal
surfaces for nanoscale patterning.
\end{abstract}

\pacs{}
\keywords{Nanostructuring, Self assembly, Scanning tunneling microscopy,
 Silicon, 1D structures}

\maketitle

Current theoretical\cite{Giamarchi2004} and technological interest in one
dimensional (1D) conductors has triggered significant efforts in fabricating
such structures.\cite{Nitzan2003,Barth2005} Conventional resist-based
lithographic methods can be extended to achieve patterning of lines on the 5-10
nm scale, using for example electron beam or nanoimprint lithography, but the
subsequent liftoff process becomes unreliable on these scales due to adhesion
failure and to line closure during development. Resist-less methods, often using
a scanning probe (SPM) to write directly on a surface, for example by plucking
single H atoms off a H-terminated Si(001) surface,\cite{Lyding1994,Hallam2007a}
can reach the level of atomic precision. But SPM lithography is not a parallel
process and thus does not readily scale to large areas or multiple structures.
The ultimate goal of assembling single-atom chains has been achieved by
single-atom manipulation using a scanning tunneling microscope
(STM).\cite{Hirjibehedin2006} However, this is a laborious process, and the
longest chain thus formed does not exceed a few tens of atoms in length.

Self-assembly, taking advantage of a material system's natural tendency to form
nanoscale structures, is an alternative to these methods with a much higher
degree of perfection and reproducibility. Surface reconstructions resulting from
metal deposition onto semiconductor surfaces can include 1D metallic chains, for
example In/Si(111)\cite{Nogami1987} and Pt/Ge(001).\cite{Gurlu2003} Metal
deposition onto vicinal Si(111) surfaces at a suitable temperature results in a
complex reconstruction which includes partially-filled metallic
states\cite{Crain2003a,Ahn2003a} located along a graphitic silicon chain that is
part of the reconstruction. On the Si(001) surface, many metals form
high-aspect-ratio epitaxial islands upon adsorption, but as discussed in a
recent review,\cite{Owen2006b} only two material systems provide 1D structures
of any great length and uniformity. Most well-known is the family of Rare-Earth
nanowires (RENW),\cite{Preinesberger1998,Owen2006b} which grow on the flat
terraces, with lengths of over 1 $\mu$m. Their shape results from anisotropic
epitaxial strain between the silicide and the silicon substrate, being almost
zero in one direction, and large in the other, with the scale and direction of
mismatch varying from one silicide to another. Recently, YSi$_{2}$ wires have
been found to exhibit characteristic behaviour of 1D structures, including van
Hove singularities and charge density waves.\cite{Zeng2008a} However, their
width varies along their length, in odd multiples of the Si lattice parameter.

Bi/Si(001) nanolines\cite{Miki1999a,Owen2002b,Owen2006b} exhibit many
appropriate features for growing self-assembled single-atom wires. Similar to
the RENW family, they grow perfectly straight on the flat terraces well clear of
any step edges, and their length, limited only by the extent of the Si(001)
terraces and by surface defects, can exceed 1 $\mu$m. But Bi nanolines offer a
decisive advantage compared to RENWs: their width is invariant, exactly 4 Si
dimers, or 1.54 nm, without any kinks.

The Bi nanolines themselves are not metallic.\cite{Owen2003} They have been used
as a template for the preferential adsorption of metals in an attempt to grow
metallic single-atom wires. Thus far, only Group III elements amongst those
surveyed have shown wire formation, with a zigzag chain
structure,\cite{Owen2006a} which, while unique and interesting, does not have
metallic properties. Hydrogenation, a standard technique for passivating the Si
surface, was used in the above templating experiments to passivate the Si, while
leaving the Bi nanolines intact. Recently, Wang et al.\cite{Wang2008a} found
that the Bi nanolines were disrupted by large doses of atomic H. Pushing this
observation to its natural extension, we exposed Bi nanolines to a H-cracker to
completely strip off the Bi from the surface. The surprising result is an
entirely new Si only 1D template formed by the Si reconstruction underneath the
Bi nanolines.

\begin{figure}
\includegraphics[width=\linewidth]{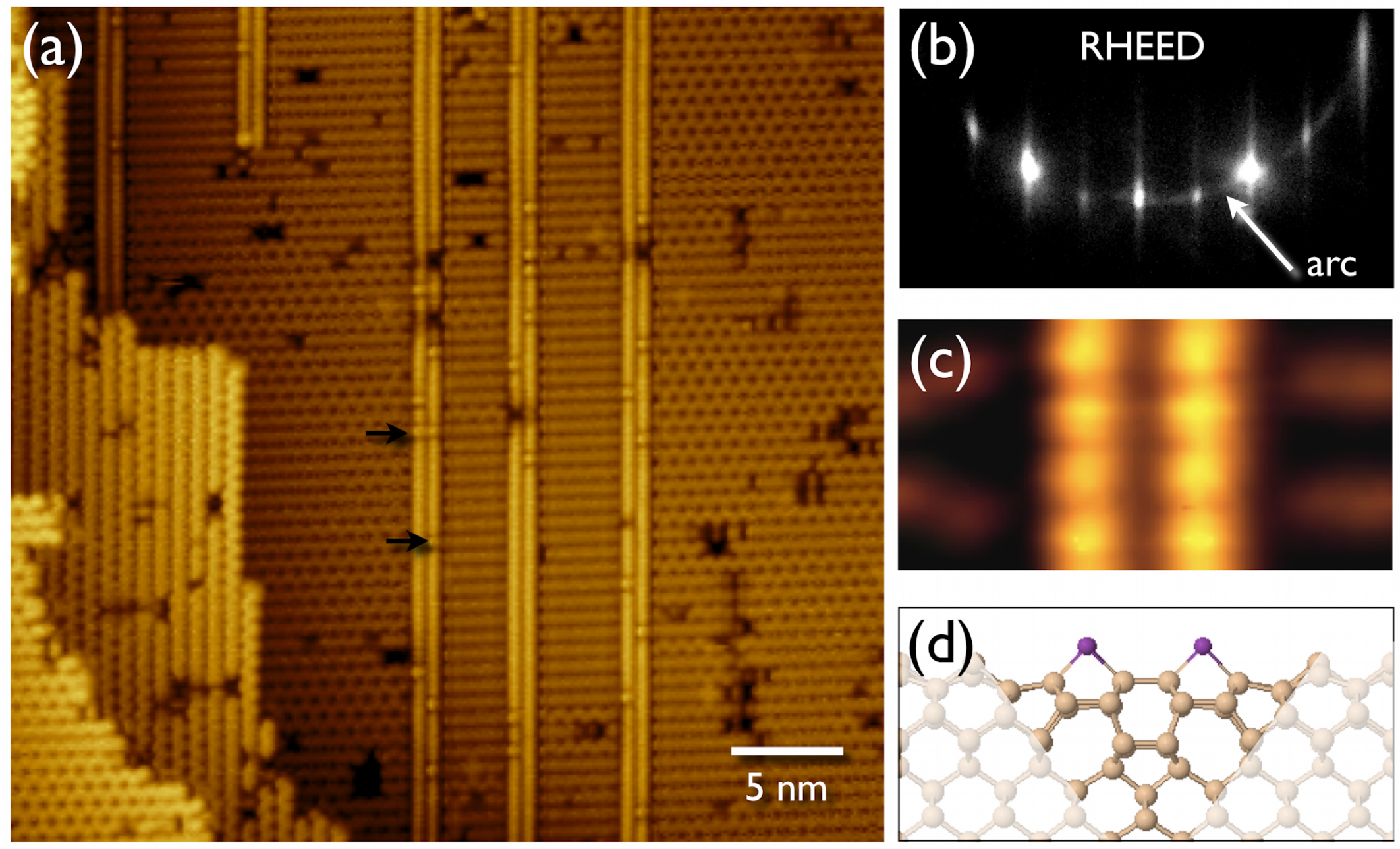}
\caption{\textbf{Bi nanolines on the clean Si(001) surface.} \textbf{a}, This
filled-states STM image ($V=-2.5$\,V, $I= 400$\,pA ) reveals the nanolines as
double bright bands perpendicular to the Si dimer rows, due to pairs of Bi
dimers. \textbf{b}, The RHEED pattern of this surface shows a characteristic
arc, caused by the Bi nanolines. \textbf{c}, High-resolution STM topographic
image resolving the Bi atoms ($V=-3.0$\,V, $I= 200$\,pA ). \textbf{d}, The Haiku
model, with two Bi dimers atop the underlying triangular reconstructed Si core.}
\label{fig1}
\end{figure}

Precursor Bi nanolines were assembled and studied in an Omicron LT-STM system.
with a base pressure in the low $10^{-11}$\,mBar range. Clean Si(001) surfaces
were prepared using a standard sequence of ex-situ etching followed by in-situ
annealing. The final in-situ step is a 15\,s flash-annealing to 1150$^{\circ}$C
with the pressure kept below $2\times10^{-9}$\,mBar. Bi was then evaporated from
a K-cell at 480$^{\circ}$C with the sample set at a temperature of approximately
570$^{\circ}$C. Typical exposure time was 13\,mins followed by an anneal at the
same temperature for 2\,mins. The progress of the nanoline growth was observed
in real-time using Reflection High-Energy Electron Diffraction (RHEED). The arc
in the RHEED pattern in Fig.~\ref{fig1}b is indicative of a well-ordered 1D
structure along [110] on the surface, in our case the Bi nanolines.

A typical STM image of Bi nanolines on the clean Si(001) surface taken at 77\,K
is shown in Fig.~\ref{fig1}a. The five bright double rows running perpendicular
to the Si(001) dimer rows are the Bi nanolines. The few dots visible on the
nanolines are most likely water adsorption onto Bi dimers, as they are not seen
in high-temperature images. Individual Bi atoms forming the bright double rows
of Bi dimers are resolved in the high resolution micrograph of Fig.~\ref{fig1}c.
The ball-and-stick cross-sectional view of the Haiku model\cite{Owen2002b} drawn
to the same scale in Fig.~\ref{fig1}d shows the registry of these Bi dimers with
the Si dimers and the structure of the underlying triangular Haiku core.

\begin{figure}
\includegraphics[width=\linewidth]{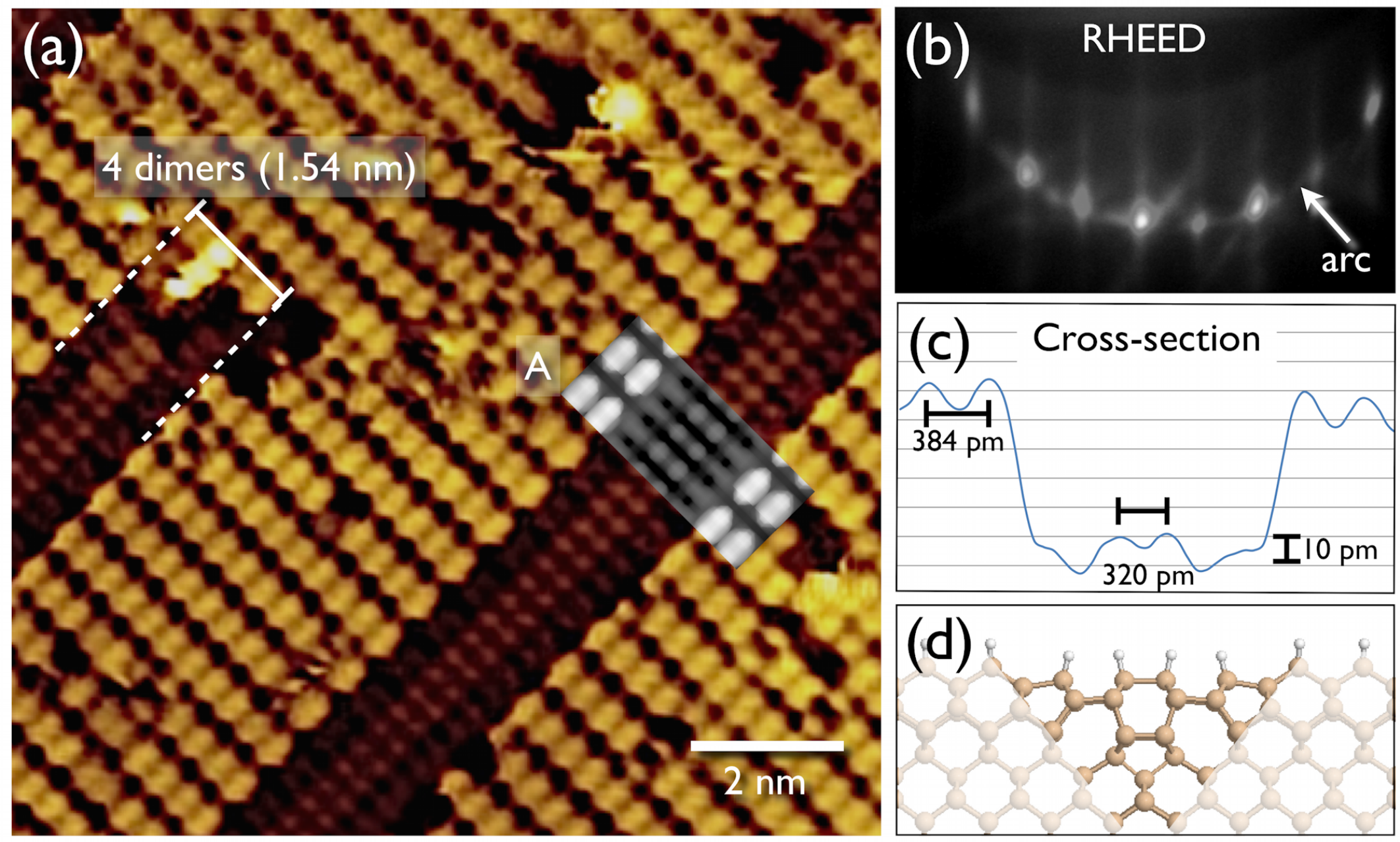}
\caption{\textbf{Haiku stripes.}
\textbf{a}, Filled-states STM image ($V=-2.8$\,V, $I= 100$\,pA ) showing two
dark stripes, exactly 4 Si dimers wide, instead of the bright double row of Bi
dimers. Two unit cells of the stripe at \textbf{A} have been replaced by a DFT
simulation. \textbf{b}, The arc in the RHEED pattern indicates that the Haiku
core of the nanolines remains intact. \textbf{c}, The cross section matches very
well the model of the Haiku core shown in panel (\textbf{d}).}
\label{fig2}
\end{figure}

We exposed Bi nanolines to atomic hydrogen using a dedicated cracker from MBE
Komponenten at 1600$^{\circ}$C and a H$_2$ pressure of $5\times10^{-7}$\,mBar.
After a 100\,s exposure with the sample set at 300$^{\circ}$C followed by a slow
cooldown to room temperature, we found that the Bi peaks had disappeared from
the X-ray Photoelectron Spectroscopy (XPS) spectra (not shown). Surprisingly,
the RHEED pattern of this hydrogenated surface clearly shows that a well-ordered
1D structure was still present (Fig.~\ref{fig2}b)

High resolution STM microscopy of the above hydrogenated surface has allowed us
to identify the 1D structure responsible for the arc seen in the RHEED pattern
after all the Bi has been removed. The image in Fig.~\ref{fig2}a was taken at a
temperature of 176 K and a pressure of $6\times10^{-12}$\,mBar. It shows that
the bright double rows of Bi dimers have been replaced by dark stripes, exactly
four Si dimers wide. Remarkably, there are no dimer-like features along them.
Instead, STM reveals a pair of well resolved dots matching each of the four Si
dimer rows spanning  the width of the stripe (i.e. 8 dots per unit cell). The
central two pairs of dots appear slightly brighter than the outer two,
corresponding to an apparent height difference in STM of about 5\,pm, as shown
in the line profile in Fig.~\ref{fig2}c. We can rule out the possibility that
the modified appearance of the nanoline is due to a voltage-dependent effect
where Bi dimers are present despite the XPS data; although Bi nanolines do
appear dark at low tunneling biases on clean Si(001) surfaces, intact Bi
nanolines never appear dark relative to hydrogenated Si(001)
surfaces.\cite{Owen2006b} Furthermore, the distinctive Bi dimer features
resolved in Fig.~\ref{fig1} are always seen when imaging Bi nanolines,
independent of STM bias.\cite{Owen2003} Thus the STM data confirms the XPS
result that the Bi dimers have been removed by the hydrogenation, and therefore
the arc seen in the RHEED pattern cannot be due to the double rows of Bi dimers
observed in Fig.~\ref{fig1}.

The simplest explanation for the atomic structure seen within the dark stripes
in Fig.~\ref{fig2} is to remove the Bi dimers from the Haiku structure, and
terminate all thus exposed Si dangling bonds with H. The precursor Bi/Si(001)
Haiku structure comprises a pair of Bi dimers which sit atop a triangular core
of reconstructed Si; this so-called Haiku core is 4 Si dimers wide and extends 5
layers below the Si(001) surface as shown in the model of Fig.~\ref{fig1}d (see
recent review\cite{Owen2006b} for further details). In the case of the stripes
the eight dots per unit cell seen by STM then correspond to two groups of four
Bi-Si backbonds of the Bi dimers, which are now Si-H bonds. A structure
generated in this way was relaxed in Density Functional Theory (DFT), and a side
view of the result is shown in Fig.~\ref{fig2}d. A plan view simulated STM image
is inset into Fig.~\ref{fig2}a (marked A), with extremely good agreement with
the experimental image. Physically, the H atoms along the stripe are 70~pm lower
than the H atoms on the background Si dimers, consistent with the height
difference of $60\pm10$ pm measured by STM. However, according to the DFT
modelling, all the H atoms on the stripe have the same physical height, and thus
the height difference between the inner and outer dots observed by STM must be
electronic in origin. DFT reveals that the HOMO state 0.3 eV below the Fermi
level has a much greater electron density at the central ring of Si in the Haiku
core. Consequently, the STM will measure a local increase in the tunnel current
at lower biases resulting in a brighter (higher) appearance in the image. The
brighter appearance of the two central atomic rows in Fig.~\ref{fig2}a is
therefore evidence of an electronic state situated at the center of the stripe.
This state could potentially support a 1D conductor, but DFT suggests that it is
not delocalised along the length of the stripe which therefore remains
non-metallic, in agreement with STM tunneling spectra (not shown).

Although the Haiku structure matches all of the experimental data available to
date on the Bi nanolines system, in particular its bias dependent STM
contrast\cite{Owen2003} it remains a contentious proposal. DFT has found it to
have the lowest energy of any Bi/Si structure calculated, including a monolayer
(ML) coverage of Bi,\cite{Owen2002b} but no direct experimental confirmation of
the reconstructed core was available to date. The pattern of dots seen by STM
within the stripe is impossible to reconcile with unreconstructed diamond Si,
but matches closely that expected from termination of the dangling bonds in the
Haiku structure by hydrogen. Hence, the hydrogenation of the Bi nanolines
provides striking evidence for the Haiku core structure.

In addition to providing the first direct experimental observation of the
existence of the Haiku structure, the Bi nanoline hydrogenation has given us an
unprecedented Si-in-Si template -- henceforth named Haiku stripe. Bi removal
affects the chemistry of the nanolines, changing their atomic wire templating
functionality. Modelling suggests that the lowest-energy sites for the
adsorption of metals such as Au\cite{Koga2007b} and
Cu\cite{Rodriguez-Prieto2009b} are by insertion into the Bi-Si bonds. Once the
Bi dimers are removed, other sites become more favourable, such as the
heptagonal rings of the Haiku core. While large kinetic barriers may prevent
some species to move subsurface (e.g. Ag) there are likely to be pathways for
others to do so at room temperature (e.g. Cu).\cite{Rodriguez-Prieto2009a}

In summary, we have presented the first direct observation of the Haiku
reconstruction and described a method for producing a novel one dimensional
Si-in-Si template that is not bound to silicon step edges. This template is
stable to about 400 $^{\circ}$C and its density can be tuned via the growth
temperature of the precursor Bi nanolines. Interestingly, DFT modelling suggests
that the heptagonal rings of the Haiku core will be stable sites for
interstitial metal adsorption, to form a new category of subsurface templated
atomic chains on Si(001). An even more significant impact of this work is that
the 1D Haiku stripe is likely to be stable in air, as it is chemically very
similar to monohydride Si(001).\cite{hersam2001} Air stability and the
subsurface adsorption sites predicted by theory, are radically expanding the
possibilities of this template. Most remarkably, we can expect atomic chains
adsorbed at subsurface sites to remain intact in air. This is a decisive
advantage over surface chains, in particular RENWs, which are very reactive and
do not survive exposure to air. Such chemical inertness holds promise for
ex-situ processing to put down contacts and thus offering the unique opportunity
for such templated nanowires to be connected to the outside world for detailed
studies of their electronic and magnetic properties as well as for inclusion
into functional devices.

\begin{acknowledgments}
This work is supported by the Swiss National Science Fundation.
The authors thank A. Rodriguez-Prieto for scientific
discussions and G. Manfrini for his expert technical support.
\end{acknowledgments}

\bibliography{haiku_stripe}

\end{document}